\documentclass[12pt,english]{article}
\usepackage[T1]{fontenc}
\usepackage[latin9]{inputenc}
\usepackage{amsmath}
\usepackage{graphicx}
\usepackage{setspace}
\usepackage{esint}
\usepackage{babel}

\begin{document}
\begin{center}
\textbf{\huge Timescape realized}
\par\end{center}{\huge \par}

\vspace{0.6cm}

\begin{center}
{\LARGE Herbert Balasin}
\par\end{center}{\LARGE \par}

\begin{center}
Institut\\
für\\
 Theoretische Physik\\
TU-Wien
\par\end{center}

\vspace{0.5cm}

\begin{center}
Wiedner Hauptstraße 8-10\\
1040 Wien\\
AUSTRIA
\par\end{center}

\vspace{1cm}

\begin{abstract}
We discuss a concrete proposal to realize the observer dependence
of expansion redshift employing a map that relates static to conformally
static geometries. This provides a manifest realization of Wiltshire's
proposal for observer selection which serves as a model to explain
accelerated expansion
\end{abstract}
\pagebreak{}

\section*{Introduction}

Recently Wiltshire \cite{Wil_1,Wil_2} has advocated a re-interpretation
of {}``dark''-energy (or in a more pristine language a cosmological
constant) as an observer selection effect purely compatible with standard
General Relativity (GR). Although his approach incorporates the general
strategy that averaging Einstein's equations may not coincide with
the Einstein-tensor of the averaged geometries -- the difference accounting
for dark energy -- an approach usually called solving the back-reaction
problem \cite{Ellis,Buch,Kor}, Wiltshire's work contains an important
additional observation connected with a genuine generally-relativistic
effect -- the non-uniqueness of time.

In particular his proposal focuses on the selection of observers due
to the fact that matter in the present epoch of the evolution of the
universe is concentrated in web-like structures that bound huge, nearly
matterless voids. His proposal relies on the fact that according to
GR regions with higher matter density slow down clocks locally, whereas
time in void-regions ticks relatively faster, neither of which needs
to coincide with cosmic time of a corresponding averaged FLRW-model.
In his original article \cite{Wil_1} Wiltshire provides an interesting
static example, namely the so-called Majumdar-Papapetrou metric \cite{MaPa},
an electro-vacuum solution of Einstein's equations consisting of an
arbitrary number of extremely-charged black-holes. On a large scale
this may be taken as constant density $k=0$ {}``Einstein-cylinder'',
whereas {}``microscopically'', within one of the bound regions,
time between two homogeneous slices in general ticks slower. As Wiltshire
himself points out this model captures only part of the idea namely
the gravitational slowing of clocks, whereas the cosmological redshift
is missing.

In the following we propose a geometrical model that incorporates
this very effect and thus may provide a simple and exact setting for
Wiltshire's idea. We will first focus on basic geometrical ideas namely
similarities between the standard gravitational redshift and its analogue
due to expansion. In the following this observation will provide a
generalization of this similarity to a map between static and conformally
static geometries, i.e. geometries admitting a timelike, hypersurface-orthogonal
Killing vector and a timelike, hypersurface-orthogonal conformal Killing
vector respectively. Thereafter we apply this setting to an arbitrary
static geometry and calculate the corresponding redshift relation
which provides an exact expression describing the mixing between cosmological
and stationary effects. This result holds in particular true for extension
of Wiltshire's original example the Majumdar-Papapetrou geometry.
However, in order to account for an initial homogeneity a simple,
rather obvious further generalization is needed. This observation
entails further that the conformal class of our models is Friedmann
or equivalently $k$-Einstein.

\section{Redshifts}

In order to motivate our approach let us consider the following simple
observation that unifies the calculation of redshift in static and
cosmological spacetimes. In order to derive the redshift in static
(stationary) geometries in a geometrical manner relative to static
(Killing-) observers \cite{Wa} one uses the geometric fact that the
inner product between the Killing $\xi^{a}$ and the tangent $k^{a}$
to an arbitrary autoparallel remains constant along the latter, i.e.\begin{eqnarray}
(k\nabla)(\xi_{a}k^{a}) & = & k^{a}k^{b}\nabla_{b}\xi_{a}=0\nonumber \\
 &  & \mbox{since }(k\nabla)k^{a}=0\mbox{ and }\nabla_{(a}\xi_{b)}=0.\label{eq:1}\end{eqnarray}
Normalizing $\xi^{a}$ turns it into the four-velocity $u^{a}$ of
the static observer and therefore the frequency of a light-signal
relative to this observer is given by\[
\omega=-u_{a}k^{a}=-\frac{\xi_{a}k^{a}}{\sqrt{-\xi^{2}}}.\]
In adapted coordinates the static geometry and it's Killing become\begin{eqnarray*}
ds^{2} & = & -V^{2}(x^{k})dt^{2}+h_{ij}(x^{k})dx^{i}dx^{j}\\
\xi^{a} & = & \partial_{t}^{a}.\end{eqnarray*}
Tagging the emitter with $E$ and a possible observer along another
Killing-trajectory with $O$, we then have\begin{eqnarray*}
\omega_{E} & = & -\left.\frac{\xi_{a}k^{a}}{\sqrt{-\xi^{2}}}\right|_{E}=-\frac{\xi_{a}k^{a}|_{O}}{\sqrt{-\xi^{2}}|_{E}}=\\
 & = & \omega_{O}\frac{\sqrt{-\xi^{2}}|_{O}}{\sqrt{-\xi^{2}}|_{E}}=\omega_{O}\frac{V(x_{O}^{k})}{V(x_{E}^{k})},\end{eqnarray*}
which is nothing but the redshift between different static observers.

Let us now consider the analogous cosmological situation. Usually
a different kind of argument is put forward relying on the spatial
symmetries of the models (cf. \cite{Wa}, p.103). However, we will
show that the situation is completely analogous to the static case
if we take into account that (\ref{eq:1}) remains intact if we replace
$\xi^{a}$ by a timelike, conformal-Killing as long as $k^{a}$ remains
null, i.e. a light-ray (which we did assume before anyways). In particular
all the FLRW-geometries admit a conformal Killing-vector $\xi^{a}$
as can be seen from\begin{eqnarray*}
ds^{2} & = & -dt^{2}+a^{2}(t)(\frac{dr^{2}}{1-kr^{2}}+r^{2}d\Omega^{2})\\
 & = & a^{2}(t)(-\frac{dt^{2}}{a^{2}(t)}+\frac{dr^{2}}{1-kr^{2}}+d\Omega^{2})\\
 & = & a^{2}(\eta)(-d\eta^{2}+\frac{dr^{2}}{1-kr^{2}}+d\Omega^{2})\\
 &  & \xi^{a}=\partial_{\eta}^{a}=a(t)\partial_{t}^{a},\end{eqnarray*}
where -- by a slight abuse of notation -- we have used the same symbol
$a$ for both the scale factor as function of $t$ as well as $\eta$
respectively. Now the redshift formula is actually completely identical,
since\begin{eqnarray*}
(k\nabla)(k^{a}\xi_{a}) & = & k^{a}k^{b}\nabla_{a}\xi_{b}=\omega k^{a}k^{b}g_{ab}\\
 &  & \mbox{since}\,\,(k\nabla)k^{a}=0,\,\, k^{a}k^{b}g_{ab}=0\,\,\mbox{and}\,\,\nabla_{(a}\xi_{b)}=\omega g_{ab}\end{eqnarray*}
 The only change with respect to the static situation comes about
from the norm of the conformal Killing vector\begin{eqnarray}
\omega_{E} & = & -u_{E}^{a}k_{a}=-\left.\frac{\xi^{a}k_{a}}{\sqrt{-\xi^{2}}}\right|_{E}=-\frac{\xi^{a}k_{a}|_{O}}{\sqrt{-\xi^{2}}|_{E}}\nonumber \\
 & = & \frac{\sqrt{-\xi^{2}}|_{O}}{\sqrt{-\xi^{2}}|_{E}}\omega_{O}=\frac{a(t_{O})}{a(t_{E})}\omega_{O}.\label{eq:5}\end{eqnarray}
Up to now these are just calculations with respect to different spacetime-families
which exhibit some similarities. However we will show that they lend
themselves to providing a map between this very spacetimes. This can
already be seen in the FLRW-context, where the $k$-{}``Einstein-cylinder''
\[
ds^{2}=-d\eta^{2}+dr^{2}/(1-kr^{2})+r^{2}d\Omega^{2}\]
 maps onto the Friedmann metric \begin{eqnarray*}
ds^{2} & = & a^{2}(\eta)(-d\eta^{2}+dr^{2}/(1-kr^{2})+r^{2}d\Omega^{2})\\
 & = & -dt^{2}+a^{2}(t)(dr^{2}/(1-kr^{2})+r^{2}d\Omega^{2})\end{eqnarray*}
 and the Killing $\partial_{\eta}^{a}$ maps to the conformal Killing
$\partial_{\eta}^{a}=a(t)\partial_{t}^{a}$ respectively.

\section{Mapping static spacetimes to spacetimes with a timelike hypersurface-orthogonal
conformal Killing vector}

Motivated by the previous analogy, we will now consider the general
situation of mapping a static spacetime $\overset{\circ}{g}_{ab}$
to a spacetime $g_{ab}$ admitting a conformal, hyperspace-orthogonal
Killing vector. Let us begin with the static spacetime, which may
be written in adapted coordinates as \begin{eqnarray*}
d\overset{\circ}{s}^{2} & = & -V^{2}(x^{k})dt^{2}+h_{ij}(x^{k})dx^{i}dx^{j}\\
 &  & \xi^{a}=\partial_{t}^{a}\qquad\xi^{2}=-V^{2},\end{eqnarray*}
which explicitly displays the Killing property $L_{\xi}\overset{\circ}{g}_{ab}=0.$
Denoting Killing-time $t$ by $\eta$ and mapping $d\overset{\circ}{s}^{2}$onto
\begin{eqnarray*}
ds^{2} & = & a^{2}(\eta)d\overset{\circ}{s}^{2}=-V^{2}(x^{k})a^{2}(\eta)d\eta^{2}+a^{2}(\eta)h_{ij}(x^{k})dx^{i}dx^{j}\\
 & = & -V^{2}(x^{k})dt^{2}+a^{2}(t)h_{ij}(x^{k})dx^{i}dx^{j},\end{eqnarray*}
we clearly see that $\partial_{\eta}^{a}=a(t)\partial_{t}^{a}$ is
now a hypersurface-orthogonal, conformal Killing of the mapped $g_{ab}=a^{2}(t)\overset{\circ}{g}_{ab}$.

Let us now turn to the converse: given a geometry $g_{ab}$ admitting
a hypersurface-orthogonal conformal Killing $\xi^{a}$ such that $\nabla_{a}\eta\in\,<\xi_{a}>$
is there a conformally related metric $\overset{\circ}{g}_{ab}$ such
that $\xi^{a}$ is a static Killing ? 

In general we have\begin{equation}
L_{\xi}g_{ab}=2\omega g_{ab}.\label{eq:2}\end{equation}
We require the existence of a conformal factor $\Omega$ such that
$\xi^{a}$ is Killing relative to the rescaled metric, i.e. \[
0=L_{\xi}(\Omega^{-2}g_{ab})=-2\Omega^{-3}L_{\xi}\Omega g_{ab}+\Omega^{-2}2\omega g_{ab},\]
which entails $L_{\xi}\Omega/\Omega=\omega$, which in turn may be
solved by a simple integration thereby establishing $\overset{\circ}{g}_{ab}=\Omega^{-2}g_{ab}$
with the required properties. More explicitly, i.e. in local coordinates
adapted to the conformal Killing vector the metric becomes\[
ds^{2}=-w(\eta,x^{k})d\eta^{2}+w_{ij}(\eta,x^{k})dx^{i}dx^{j}\qquad\xi^{a}=\partial_{\eta}^{a}.\]
Condition (\ref{eq:2}) is turned into\begin{equation}
-w'd\eta^{2}+w'_{ij}dx^{i}dx^{j}=-2\omega wd\eta^{2}+2\omega w_{ij}dx^{i}dx^{j},\label{eq:3}\end{equation}
where the prime denotes the derivative with respect to $\eta$. We
thus obtain from (\ref{eq:3}) two conditions, the first of which
$(logw)'=2\omega$ is readily integrated to give \[
w(\eta,x^{k})=\overset{\circ}{w}(x^{k})e^{2\int\limits _{\eta_{0}}^{\eta}d\eta'\omega(\eta')}\]
and the second becomes $w_{ij}'=(logw)'w_{ij}$ and thus\[
w_{ij}(\eta,x^{k})=\overset{\circ}{w}_{ij}(x^{k})w(\eta,x^{k})=\overset{\circ}{w}(x^{k})\overset{\circ}{w}_{ij}(x^{k})e^{2\int\limits _{\eta_{0}}^{\eta}d\eta'\omega(\eta')}.\]
Therefore we find for the static counterpart $\overset{\circ}{g}_{ab}$
of the metric $g_{ab}$\begin{eqnarray}
ds^{2} & = & -\overset{\circ}{w}e^{2\int\limits _{\eta_{0}}^{\eta}d\eta'\omega(\eta')}d\eta^{2}+\overset{\circ}{w}\overset{\circ}{w}_{ij}e^{2\int\limits _{\eta_{0}}^{\eta}d\eta'\omega(\eta')}dx^{i}dx^{j}\nonumber \\
 & = & -\overset{\circ}{w}dt^{2}+a^{2}(t)w_{ij}dx^{i}dx^{j},\label{eq:4}\\
d\overset{\circ}{s}^{2} & = & -\overset{\circ}{w}(x^{k})d\eta^{2}+w_{ij}(x^{k})dx^{i}dx^{j},\nonumber \end{eqnarray}
where we used $dt=e^{\int\limits _{\eta_{0}}^{\eta}\omega(\eta')}d\eta$.
This is precisely the form that we started from\[
g_{ab}=a^{2}(\eta)\overset{\circ}{g}_{ab}\]
 which shows that our map between geometries is invertible.

\section{Redshift for geometries with a static, conformal Killing vector}

Let us now consider the change in the redshift formula for our mapped
geometry. Starting from \begin{equation}
ds^{2}=-w(x^{k})dt^{2}+a^{2}(t)w_{ij}(x^{k})dx^{i}dx^{j}\label{eq:6}\end{equation}
with conformal Killing vector $\xi^{a}=a(t)\partial_{t}^{a}$ we obtain
for the frequencies of two observers following conformal Killing trajectories
\begin{equation}
\omega_{E}=\frac{\sqrt{-\xi^{2}}|_{O}}{\sqrt{-\xi^{2}|_{E}}}\omega_{O}=\left(\xi^{2}=-a^{2}(t)w(x^{k})\right)=\frac{a(t_{O})\sqrt{w(x_{O}^{k})}}{a(t_{E})\sqrt{w(x_{E}^{k})}}\omega_{O}.\label{eq:7}\end{equation}
 In particular for the Majumdar-Papapetrou geometry\[
d\overset{\circ}{s}^{2}=-\frac{1}{V^{2}}dt^{2}+V^{2}dx^{i}dx^{i}\qquad V=1+\sum_{i}\frac{2M_{i}}{|x^{m}-x_{i}^{m}|}\]
we obtain\begin{equation}
\omega_{E}=\frac{a(t_{O})}{V(x_{O}^{k})}\frac{V(x_{E}^{k})}{a(t_{E})}\omega_{O},\label{eq:8-1}\end{equation}
which depends now on spatial position as well as time.

These simple relations clearly display Wiltshire's idea that gravitational
redshift is modified by the observer position. Only in an averaged
sense does the original FLRW-formula (\ref{eq:5}) hold, i.e. if $V$
(or $w$) becomes constant.

\section{Einstein tensor for geometries with a static, conformal Killing vector}

Up to now our considerations have been purely kinematical. In order
to incorporate dynamics we have to calculate the Einstein-tensor for
conformally related geometries. The metric and difference tensor between
the corresponding Levi-Civita derivatives are given by\begin{eqnarray*}
g_{ab} & = & \Omega^{2}\overset{\circ}{g}_{ab}\\
C^{a}\,_{bc} & = & \overset{\circ}{\nabla}_{b}log\Omega\,\delta_{c}^{a}+\overset{\circ}{\nabla}_{c}log\Omega\,\delta_{b}^{a}-\overset{\circ}{\nabla}\,^{a}log\Omega\,\overset{\circ}{g}_{bc}\end{eqnarray*}
From this and the expression of the Riemann-tensor \[
R^{a}\,_{bcd}=\overset{\circ}{R}\,^{a}\,_{bcd}+\overset{\circ}{\nabla}_{c}C^{a}\,_{bd}-\overset{\circ}{\nabla}_{d}C^{a}\,_{bc}+C^{a}\,_{mc}C^{m}\,_{bd}-C^{a}\,_{md}C^{m}\,_{bc}\]
we obtain the Einstein-tensor\begin{eqnarray}
G_{ab}=R_{ab}-\frac{1}{2}Rg_{ab} & = & \overset{\circ}{R}_{ab}-\frac{1}{2}\overset{\circ}{g}_{ab}\overset{\circ}{R}-2\overset{\circ}{\nabla}_{a}\overset{\circ}{\nabla}_{b}log\Omega+2\overset{\circ}{g}_{ab}\overset{\circ}{\nabla}\,^{2}log\Omega\nonumber \\
 &  & +2\overset{\circ}{\nabla}_{a}log\Omega\overset{\circ}{\nabla}_{b}log\Omega+\overset{\circ}{\nabla}_{m}log\Omega\overset{\circ}{\nabla}\,^{m}log\Omega\overset{\circ}{g}_{ab}\label{eq:8}\end{eqnarray}
In our particular case $\Omega=a$ and for the metric (\ref{eq:6})
the above becomes

\begin{eqnarray*}
R_{ab}-\frac{1}{2}Rg_{ab} & = & \overset{\circ}{R}_{ab}-\frac{1}{2}\overset{\circ}{g}_{ab}\overset{\circ}{R}-2\left(\frac{a'}{a}\right)^{'}\frac{1}{w^{2}}\overset{\circ}{\xi}_{a}\overset{\circ}{\xi}_{b}-\left(\frac{a'}{a}\right)\frac{2}{w^{2}}\overset{\circ}{\nabla}_{(a}w\overset{\circ}{\xi}_{b)}\\
 &  & -\frac{2}{w}\left(\frac{a'}{a}\right)^{'}\overset{\circ}{g}_{ab}+2\left(\frac{a'}{a}\right)^{2}\frac{1}{w^{2}}\overset{\circ}{\xi}_{a}\overset{\circ}{\xi}_{b}-\frac{1}{w}\left(\frac{a'}{a}\right)^{2}\overset{\circ}{g}_{ab}\end{eqnarray*}
which display a structure similar to the FLRW-models due to the appearance
of the containing $\dot{a}/a$ and its derivative. The static Einstein-tensor
plays the role of the spatial curvature. This is actually no co-incidence
since for the {}``Einstein-cylinder'' as static geometry our transformation
yields the corresponding FLRW-expression.

\section{A possible generalization - conformal Friedmann}

Although the previous sections show the main features of the timescape
proposal a closer look at the derivation of the map between static
geometries and geometries admitting a timelike, hypersurface-orthogonal,
conformal Killing vector-field allows a rather natural generalization.
To this end let us re-consider the derivation of the map and begin
with an arbitrary geometry admitting a timelike, hypersurface-orthogonal,
conformal Killing vector-field $\xi^{a}.$In adapted coordinates $\xi^{a}$
and the metric $g_{ab}$ become\begin{eqnarray*}
ds^{2} & = & -w(x^{k},\eta)d\eta^{2}+w_{ij}(x^{k},\eta)dx^{i}dx^{j}\\
\xi^{a} & = & \partial_{\eta}^{a}\end{eqnarray*}
 which yields via the conformal Killing equation $L_{\xi}g_{ab}=2\omega g_{ab}$

\begin{eqnarray*}
w'(x^{k},\eta) & = & 2\omega(x^{k},\eta)w(x^{k},\eta)\\
w'_{ij}(x^{k},\eta) & = & 2\omega(x^{k},\eta)w_{ij}(x^{k},\eta).\end{eqnarray*}
These relations are solved by\begin{eqnarray*}
w(x^{k},\eta) & = & \overset{\circ}{w}(x^{k})e^{2\int\limits _{\eta_{0}}^{\eta}\omega(x^{k},\eta')d\eta'},\\
w'_{ij}(x^{k},\eta) & = & \frac{w'(x^{k},\eta)}{w(x^{k},\eta)}w_{ij}(x^{k},\eta),\\
w_{ij}(x^{k},\eta) & = & w(x^{k},\eta)\overset{\circ}{w}_{ij}(x^{k}).\end{eqnarray*}
Therefore we have\[
ds^{2}=w(x^{k},\eta)(-d\eta^{2}+\overset{\circ}{w}_{ij}(x^{k})dx^{i}dx^{j})\]
or if we denote $w(x^{k},\eta)=a^{2}(\eta,x^{k})$\begin{equation}
g_{ab}=a^{2}(\eta,x^{k})\overset{\circ}{g}_{ab}\label{eq:9}\end{equation}
where $\overset{\circ}{g}_{ab}=-d\eta_{a}d\eta_{b}+\overset{\circ}{w}_{ij}(x^{k})dx_{a}^{i}dx_{b}^{j}$
is the corresponding static geometry with $\xi^{a}=\partial_{\eta}^{a}$
as Killing-vector%
\footnote{Actually the last gives rise to an even simpler, geometrical proof.
Namely taking the conformal Killing norm we have $\overset{\circ}{g}_{ab}=g_{ab}/(-\xi^{2})$
and if we take $(\xi\nabla)\xi^{2}=1/2(\nabla\xi)\xi^{2}$ into account
$L_{\xi}\overset{\circ}{g}_{ab}=0$ follows.%
}. Moreover, relative to the static geometry $\xi^{a}$ is covariantly
constant, since $\nabla_{a}\overset{\circ}{\xi}_{b}=\nabla_{[a}\overset{\circ}{\xi}_{b]}+\nabla_{(a}\overset{\circ}{\xi}_{b)}$
and the first term vanishes due to exactness of $\overset{\circ}{\xi}_{a}=-d\eta_{a}$
and the second by $\xi^{a}$ being Killing. At first glance this might
seem strange since in the previous section the static Killing was
not covariantly constant. However the previous results may easily
be recovered as the special case of factorizing conformal factor,
i.e. $a^{2}(\eta,x^{k})=a^{2}(\eta)w(x^{k})$, which allows to bring
the metric into the {}``preferred'' form\begin{eqnarray*}
ds^{2} & = & a^{2}(\eta,x^{k})(-d\eta^{2}+\overset{\circ}{w}_{ij}(x^{k})dx^{i}dx^{j})\\
 & = & a^{2}(\eta)(-w(x^{k})d\eta^{2}+w_{ij}(x^{k})dx^{i}dx^{j})\\
 & = & -w(x^{k})dt^{2}+a(t)w_{ij}(x^{k})dx^{i}dx^{j}.\end{eqnarray*}
This is precisely the result obtained earlier. However, by admitting
arbitrary (i.e. non-factorizing) conformal factors the static part
is only defined up to a static, (i.e. $x^{k}$-dependent) conformal
transformation.

The redshift relation (\ref{eq:8-1}) becomes\begin{equation}
\omega_{E}=\frac{a(\eta_{O},x_{O}^{k})}{a(\eta_{E},x_{E}^{k})}\omega_{O.}\label{eq:10}\end{equation}
Taking the emission time $\eta_{E}$ to be (close to) decoupling and
the corresponding conformal factor position-independent, we easily
see that the redshift expression \[
\omega_{E}=\frac{a(\eta_{O},x_{O}^{k})}{a(\eta_{E})}\omega_{O}\]
depends only on the observer position. This is actually an important
pre-requisite that would have been impossible with a factorizing conformal
factor. In particular each {}``conformal'' observer sees an isotropic
red-shift but with different temperatures depending on the observer
location, as pointed out by Wiltshire.\\

\includegraphics[scale=0.5]{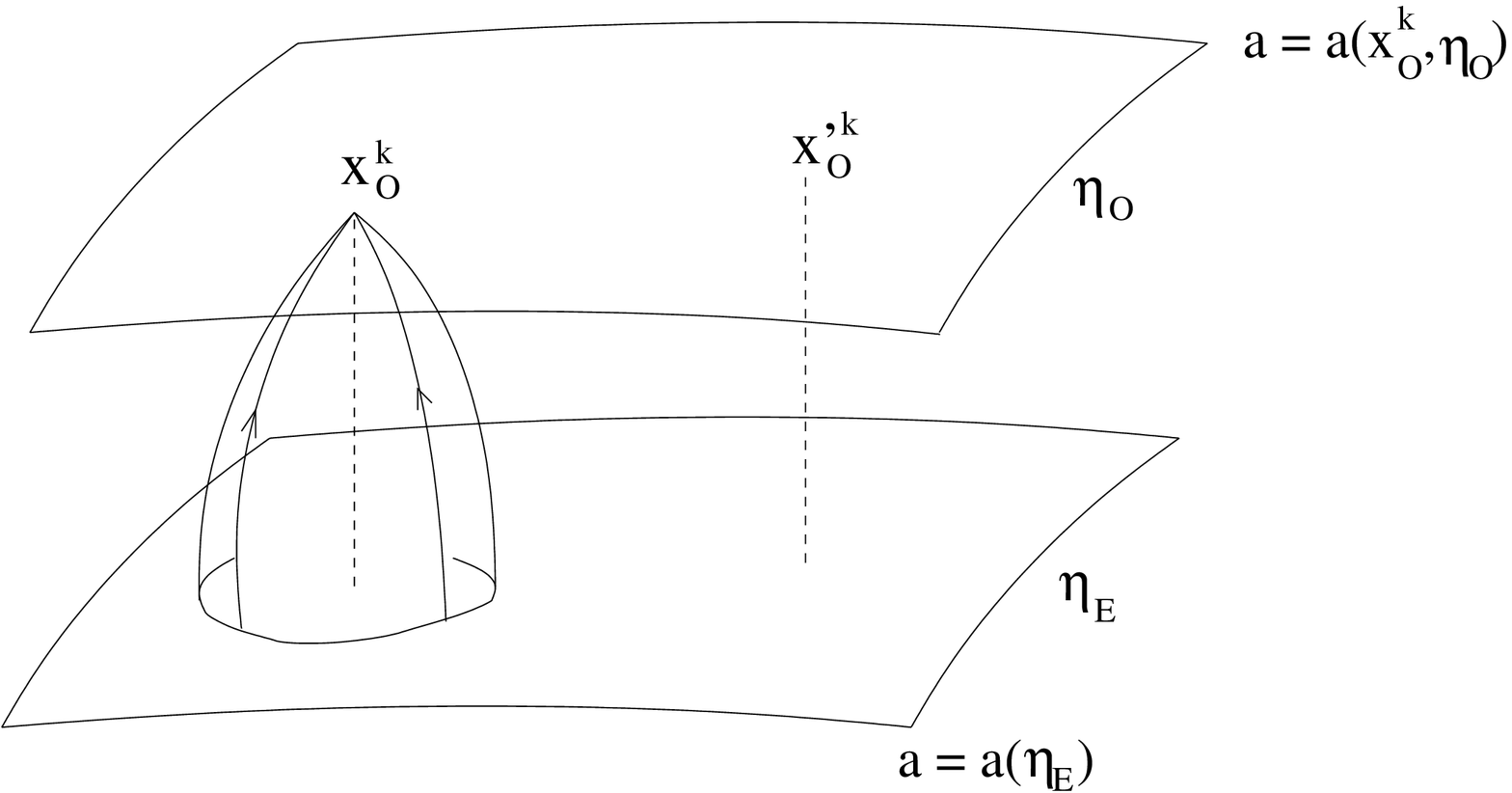}\\
\\
Finally the Einstein tensor with respect to (\ref{eq:8}) and (\ref{eq:9})
becomes \begin{eqnarray*}
G_{ab} & = & \overset{\circ}{\xi}_{a}\overset{\circ}{\xi}_{b}(\frac{1}{2}\,^{3}\overset{\circ}{R}+3\left(\frac{a'}{a}\right)^{2}-2\overset{\circ}{D}\,^{2}\log a-\overset{\circ}{D}_{m}\log a\overset{\circ}{D}\,^{m}\log a)\\
 &  & +2\overset{\circ}{\xi}_{a}(\overset{\circ}{D_{b}}(\frac{a'}{a})-(\frac{a'}{a})\overset{\circ}{D}_{b}\log a)+2\overset{\circ}{\xi}_{b}(\overset{\circ}{D_{a}}(\frac{a'}{a})-(\frac{a'}{a})\overset{\circ}{D}_{a}\log a)\\
 &  & +\left[\,^{3}\overset{\circ}{G}_{ab}-2\overset{\circ}{D}_{a}\overset{\circ}{D}_{b}\log a+2\overset{\circ}{D}_{a}\log a\overset{\circ}{D}_{b}\log a+\right.\\
 &  & +\left.\overset{\circ}{w}_{ab}(-2(\frac{a'}{a})'-(\frac{a'}{a})^{2}+2\overset{\circ}{D}\,^{2}\log a+\overset{\circ}{D}_{m}\log a\overset{\circ}{D}\,^{m}\log a)\right].\end{eqnarray*}
In particular the terms proportional to the tensor-square of $\overset{\circ}{\xi}_{a}$
exhibit an even greater similarity to the Friedmann-equation than
in the case of factorizing conformal factor. This is mainly due to
the fact that corresponding static geometry in the general case is
closer to the $k$-{}``Einstein-cylinder''. 

In order to provide a model for the right-hand side of the Einstein-equations
we consider coupling to standard, i.e. matter-dominated perfect-fluids
\[
G_{ab}=8\pi T_{ab}\qquad T_{ab}=\rho u_{a}u_{b.}\]
Covariant conservation of $T_{ab}$, i.e. $\nabla^{a}T_{ab}=0$, which
follows from the Einstein-equations, requires\[
(u\nabla)u_{b}\rho+\nabla_{a}(\rho u^{a})u_{b}=0,\]
which in particular entails the geodeticity of matter, i.e. $(u\nabla)u^{a}=0.$
However, for the observer comoving with the expansion, i.e. the conformal
Killing $\xi^{a}$ has $\hat{u}^{a}=\xi^{a}/\sqrt{-\xi^{2}}$ . Therefore\[
(\hat{u}\nabla)\hat{u}^{a}=\frac{1}{-\xi^{2}}(\xi\nabla)\xi^{a}+\frac{\xi^{a}}{(-\xi^{2})^{2}}\xi_{c}(\xi\nabla)\xi^{c}=\frac{1}{-\xi^{2}}h^{a}\,_{c}(\xi\nabla)\xi^{c}.\]
The conformal Killing-property of $\xi^{a}$ entails\begin{eqnarray*}
(\xi\nabla)\xi_{a} & = & -\xi^{c}\nabla_{a}\xi_{c}+\frac{1}{2}(\nabla\xi)\xi_{a}\\
 & = & -\frac{1}{2}\nabla_{a}\xi^{2}+\frac{1}{2}(\nabla\xi)\xi_{a},\end{eqnarray*}
and therefore the acceleration becomes\[
(\hat{u}\nabla)\hat{u}^{a}=\frac{1}{2\xi^{2}}h^{ac}\nabla_{c}\xi^{2},\]
which shows that the latter is in general non-vanishing if the conformal
factor has a spatial variation as required by the timescape model.
Therefore we find a relative velocity between the matter and the expansion
frame as seems to be suggested by observations like {}``dark flow''
\cite{Kash,Wil_3} .

\section*{Summary and Conclusion}

In the present note we put forward, based on a simple geometrical
observation (from the redshift relations) that static and cosmological
spacetimes can actually be treated on the same footing. More precisely
this idea gives rise to a map between these geometries. In particular
this map should provide a concrete realization of Wiltshire's proposal
of observer dependence for the redshift as can be seen from (\ref{eq:7}),(\ref{eq:10}).
Taking a closer look at the expression we see that it becomes the
standard FLRW-relation if we consider an averaged, i.e. spatially-constant,
homogeneous Killing norm. Although our proposal was originally motivated
by a {}``factorizing'' conformal factor, as in (\ref{eq:6}), it
is important to note that our results hold even in the more general
case discussed in the previous section. This generalization exhibits
the possibility of an isotropic, albeit position-dependent, red-shift.
Moreover the local conformal relation $ds^{2}=\Omega^{2}(-d\eta^{2}+d\sigma_{k}^{2})$
shows that the expansion vs. proper time relation is position independent,
i.e. $\Delta\tau=\Omega\Delta\eta$and $\Delta l=\Omega\Delta\sigma$
imply $\Delta l/\Delta\tau=\Delta\sigma/\Delta\eta$. It is clear
that our observation is just a first preliminary step, but we do hope
that it provides a starting point for geometries that can be turned
in homogeneous and isotropic models upon averaging and still provides
some of the {}``microscopic'' effects that should account for accelerated
expansion in Wiltshire's timescape proposal.

\end{document}